# COMMISSIONING AND FIRST RESULTS FROM CHANNELING RADIATION AT FAST*

A. Halavanau†, D. Mihalcea, (NIU, DeKalb, IL) D. Broemmelsiek, D. Edstrom Jr.,
T. Sen, A. Romanov, J. Ruan, V. Shiltsev (FNAL, Batavia, IL) P. Kobak (BYU-I, Rexburg,
ID), W. Rush (KU, Lawrence, KS, USA), J. Hyun (Sokendai, Ibaraki, Japan)

*Abstract*

X-rays have widespread applications in science and industry, but developing a simple, compact, and high-quality X-ray source remains a challenge. Our collaboration has explored the possible use of channeling radiation driven by a 50 MeV low-emittance electron beam to produce narrow-band hard X-rays with photon energy of 40 to 140 keV [8,9,11]. Here we present the simulated X-ray spectra including the background bremsstrahlung contribution, and a description of the required optimization of the relevant electron-beam parameters necessary to maximize brilliance of the resulting X-ray beam. Results are presented from our test of this, carried out at the Fermilab Accelerator Science & Technology (FAST) facility's 50-MeV low-energy electron injector. As a result of the beam parameters, made possible by the photo-injector based SRF linac, the average brilliance at FAST was expected to be about one order of magnitude higher than that in previous experiments.

## INTRODUCTION

Crystal channeling presents the possibility of developing an easily disseminable, compact, and high-quality X-Ray source, which could be useful in a number of areas of science and industry (e.g. lithography and adhesive curing). Because of the conditions for crystal channeling, a low-emittance electron beam is necessary for significant X-ray production. The Fermilab Accelerator Science & Technology (FAST) facility's photoinjector-based SRF linac is ideal for testing this. Details of the UV Drive laser, photocathode-based electron gun, and other current machine parameters have been noted elsewhere [1].

The crystal channeling test setup consisted of a diamond crystal mounted in a goniometer (provided by HZDR via Vanderbilt University, see Fig. 1) near the end of the low-energy linac section tested during the 50-MeV commissioning run this past summer (2016). An open channel or Al foil target may also be used, selected with the goniometer translation axis. Once a channel is found by adjusting the pitch and yaw axes of the diamond crystal with the goniometer, the electrons oscillating within the crystal channel generate a co-linear X-ray beam. The electrons were then swept downward into the low-energy beam absorber, but the X-rays were allowed to pass through a diamond window at the end of the low energy beamline for detection.

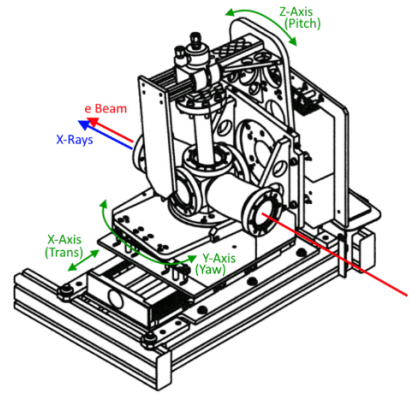

Figure 1: The goniometer from Helmholtz Zentrum Dresden Rossendorf (HZDR) used in the crystal channeling efforts during the 50-MeV run [10].

*Detector Commissioning*

To observe channeling two detectors were proposed: a forward or primary-beam detector (FwD) and a Compton-scattering detector (CSD) as seen in Figure 2. Both single-photon Amptek X-ray spectrometers, the FwD was located roughly 2 m from a channeling crystal. Because of the sensitivity of the detectors, a piece of PVC was placed in the path of the expected X-ray trajectory at 45° roughly 1.5 m from the diamond crystal to serve as a Compton scattering surface with the CSD 1 m from that to detect the much lower-intensity scattered X-ray beam expected from the scattering surface (7 orders of magnitude below the primary beam), allowing the CSD to operate at higher electron bunch charges. Installation and alignment of both detectors has been described previously [2]. Specifications for the detectors are listed in Table 1.

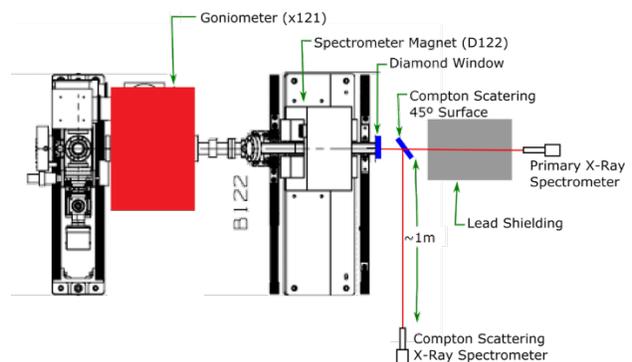

Figure 2: Forward and 90-deg (Compton scattering) detector configuration.

Both detectors were calibrated with $^{57}$Co source, which produces emission at 14.4 keV and 122 keV. This tests much of the desired energy range for the crystal channeling experiment as seen in Figure 3. For the purpose of the experiment, the spectrometer lower energy cut-off was set to 15 keV to cut out noise from the bremsstrahlung (BS) below the peaks of interest expected from crystal channeling.

Spectra were acquired using python-based acquisition software coordinated through the Fermilab controls native Accelerator Command Language (ACL) and python routines [3]. Acquisition times were set through the coordination routines to allow for some flexibility in spectra collection. With the typical machine cycle rate of 1 Hz, the typical acquisition time allowed for integration over 300 machine cycles, but some longer integration periods (up to 1 hour) were used for some tasks.

Table 1: FwD and CSD specifications.

| X-Ray Spectrometer | Amptek CdTe X-123 |
|---|---|
| Sensitive Area | 9 mm$^2$ |
| Selected Energy Range | 15 keV – 150 keV |
| Number of Channels | 1024 |

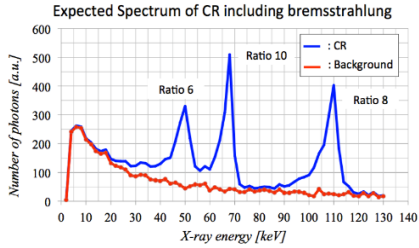

Figure 3: A simulated crystal channeling spectrum. Note that this is for 20 pC/pulse electrons at 43 MeV in the crystal channel, but does not take dark current into account.

*Linac commissioning*

As specified elsewhere [1], quadrupole scans were used to perform emittance measurements near the location of the diamond crystal, determination of the beam Twiss parameters there being a natural by-product of this measurement. Final alignment of the beam down the line was performed with beam-based techniques to center the beam through the RF cavities and beamline quadrupoles thereby minimizing the dipole components with respect to phases and amplitudes respectively. The RF centering relied upon conjugate gradient minimization while the quadrupole centering was performed with an iterative, thick-element model-dependent technique [1,4].

Initially the quadrupoles in the 50 MeV beamline were significantly stronger, resulting in a harder focus at the diamond crystal, but these were relaxed near the end of the commissioning run. This was done to minimize the angular span through the crystal allowing more of the beam to find a channel so long as the crystal orientation was correct.

While beam currents from the gun have been run as high as 4 nC/pulse, projected saturation levels for the Amptek detectors required significantly less. The nominal configuration required a neutral density filter to be placed in the photoinjector UV drive laser path, which limited maximum production to 200 pC/pulse (still detectable by BPMs and the beamline toroids) at full transmission to the PC.

Even at 200 pC/pulse the radiation flux from channeling would be several orders of magnitudes higher than the maximum capability of the detectors linear operation limit. Studies were performed to determine and understand the detector performance. The injector optics were re-optimized for low charge (≤50 fC/pulse). In order to produce these bunch charges, several adjustments to the laser system control were made as described elsewhere. [1]

To examine detector response, the counts from each detector were integrated for 300 s at a number of charge steps (see Fig. 4). The FwD confirmed linear scaling up to charges of 50 fC/pulse and the CSD preserves the linearity feature up to 10 pC/pulse. These injector settings were selected for the later experiments.

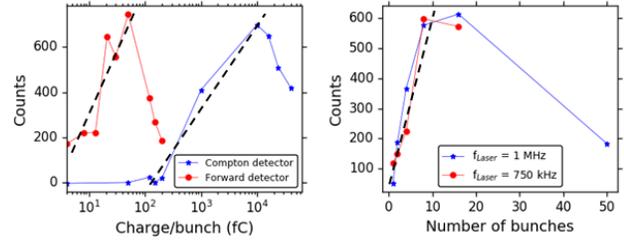

Figure 4: Scaling of Forward and 90-deg detectors counts vs. bunch charge (left) and number of counts in 90-deg detector as a function of number of bunches (right).

*Dark current*

Dark current was a significant issue and its mitigation was not trivial. A number of approaches were taken in parallel, including optimization of the gun, collimation, and energy-scraping through the chicane.

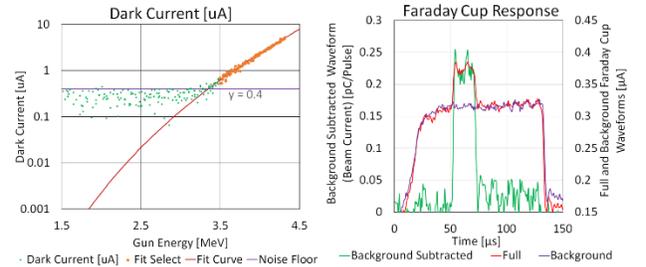

Figure 5: The gun dark current was reduced significantly by reducing the gun exit energy to ~3.5 MeV (left). Even with the lower gun gradient, dark current from the gun dominated the overall bunch signal, as shown (right) for a 200 fC/pulse signal.

Optimization of the gun involved both lowering the cavity gradient and shortening the RF pulse to the cavity to more closely match the number of pulses in the pulse train provided by the drive laser within the macropulse (see Fig. 5). Reducing the gradient, and consequently the RF-gun field, however also reduces the overall beam energy resulting in a trade-off with space-charge effects and greater overall focusing through the SRF structures that follow.

The latter generally results in higher emittance and consequently channeling is impaired.

Between the electron gun and the first SRF cavity, a collimator can be inserted and was generally used when not checking beam intensity with the Faraday cup as they occupy the same space in the beamline when inserted. The collimator was intended to be used in conjunction with a dark current kicker to provide longitudinal scraping of dark current as well, but the kicker has not been installed, limiting the reduction in dark current only to the transverse collimator aperture [5]. The gun cavity solenoids were adjusted to optimally scrape remaining dark current.

Throughout the crystal collimation effort, beam was accelerated through the two SRF booster cavities that follow the gun to ~43 MeV. Because the beam itself was accelerated on-crest through CC1 (maximum energy gain), it follows that any dark current remaining will have lower energy. The chicane being a dispersive segment of the low energy beamline, allowed selective scraping of the lower-energy dark current.

The amount scraped away with the crystal collimation was not well regulated because the air-cooled chicane dipoles (ChDs) are not locked to an NMR probe (as the spectrometer magnet, D122, is). This results in field fluctuations that can easily be exacerbated by beamline enclosure conditions (temperature fluctuation, air currents, etc). Nominally the amount of beam scraped away for these studies ranged from 10% to 50% depending on the other specific conditions of the study. This, however, may have contributed to the background seen by the CSD, as it was in the same plane and as the chicane and likely lacked adequate shielding.

## RESULTS

The first test of the FwD performance was made with the Al foil goniometer target (see Fig. 6). The total exposure time of 30 min corresponds to the spectrum shown in Figure 6. It is in agreement with Kramers' BS formula [6].

GEANT4 [7] simulations (Fig. 7) suggest this BS spectra is in fact a superposition of the various BS spectra for the various constituent beamline components. In particular, this is true for detecting the second channeling spectral line, at 72 keV, due to the stainless steel BS signature. Also, note that several Pb K-lines lay around 75 keV.

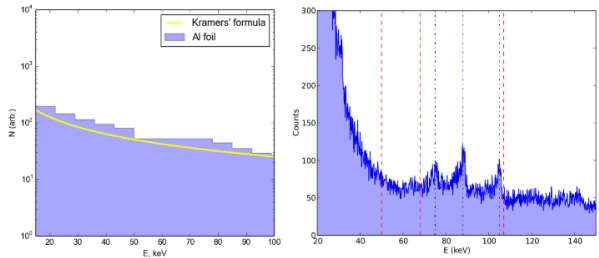

Figure 6: BS spectrum of the Al foil and it's fit with Kramers's formula (right) and the sum of measured spectrua for a 30 fC/pulse electron beam. Red dashed lines correspond to predicted channeling lines.

A total of 250 30-fC/pulse channeling spectra were summed together for electron beam. While it was not obvious at the time of data collection, three peaks appear over the BS background spectrum (see Fig. 6). The discrepancy between this and the simulation (Fig. 3) is a subject of an ongoing investigation and may be due to nonlinear detector performance in radiation environment, crystal morphological changes due to aging, or beamline energy measurement offset (unmeasurable during crystal collimation due to intensity limitations).

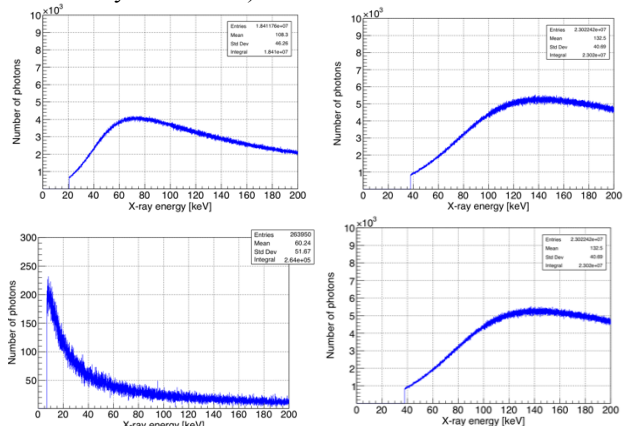

Figure 7: GEANT4 simulations of BS spectra of the 4 most common FAST beamline materials: 2 mm stainless steel (top left), 5 mm Cu frame (top right), 0.5 mm Al foil (low left) and 3 mm Nb (low right).

## CONCLUSION

Crystal channeling was investigated in the FAST 50-MeV run in 2016. A number of specific beam and dark-current related issues were resolved and optics solutions were found to generate and deliver ultra-low charge (<100 fC/pulse) electron beam to a diamond crystal.

Dark current mitigating solutions included optimization of the gun and propagation of the electron beam through the chicane, which may resulted in a significantly higher background. A low intensity X-ray signal from Al foil was successfully acquired by the FwD and it corresponds to analytical prediction. Significant X-ray radiation (e.g. from channeling radiation) causes saturation and detector pile-up. The full impact of this process on data collection is the subject of ongoing study. A number of possible solutions are being considered in the hopes of continuing work on crystal channeling in the future at FAST.

## ACKNOWLEDGEMENT

We acknowledge technical support from the FAST team, including members of the Accelerator Division and Technical Division, Dr. Philippe Piot from NIU/FNAL for enormous contribution to this experiment, Dr. Wolfgan Wagner formerly from Helmholtz-Zentrum Dresden-Rossendorf (HZDR) Dresden, Germany (now at MePhy, Moscow, Russia) for providing the goniometer.